\documentclass[10pt]{article}

\usepackage{amsmath}
\usepackage{amssymb}
\usepackage{hyperref}
\usepackage{lineno}
\usepackage{longtable}
\usepackage{graphicx}

\title{Computational models of consumer confidence from large-scale online attention data: crowd-sourcing econometrics.}
\author{Xianlei Dong$^{1,2,\ast,\P}$,  Johan Bollen$^{2,\P}$}

\begin{document}

\maketitle

\begin{flushleft}
\textbf{1} School of Economics and Management, Beijing University of Technology, Beijing, China\\
\textbf{2} School of Informatics and Computing,   Indiana University, United States of America\\

$\ast$ Corresponding author: sddongxianlei@163.com (XD)\\
\P: These authors contributed equally to the manuscript.
\end{flushleft}	

\section*{Abstract}
Economies are instances of complex socio-technical systems that are shaped by the interactions of large numbers of individuals. The individual behavior and decision-making of consumer agents is determined by complex psychological dynamics that include their own assessment of present and future economic conditions as well as those of others, potentially leading to feedback loops that affect the macroscopic state of the economic system. We propose that the large-scale interactions of a nation's citizens with its online resources can reveal the complex dynamics of their collective psychology, including their assessment of future system states. Here we introduce a behavioral index of Chinese Consumer Confidence (C3I) that computationally relates large-scale online search behavior recorded by Google Trends data to the macroscopic variable of consumer confidence. Our results indicate that such computational indices may reveal the components and complex dynamics of consumer psychology as a collective socio-economic phenomenon, potentially leading to improved and more refined economic forecasting.

\section*{Introduction}

The growth of most modern economies is driven by consumer spending \cite{Deaton1980}. Therefore, consumer confidence levels can 
have significant effects on economic growth. Consumer Confidence Indices (CCI) are designed to measure the degree of
confidence that consumers have with respect to the state of the economic system. The basis for many CCIs lies in behavioral science
where evidence has accumulated that individual consumer behavior is influenced by a number of
emotional and social factors \cite{Kietzmann2011,Bollen2011} that interact with the consumer agents' socio-economic context.
In other words, the emotional state of consumers as well as their 
assessment of that of other consumers will shape their subsequent individual consumption patterns \cite{Frijda1994, Shi2014}.
In the aggregate, as consumers collectively lose or gain confidence in the state of the economy, this is assumed to affect their
collective consumption patterns and thus economic growth yielding a complex interaction between consumer
confidence and economic conditions. This interplay between the complex behavior of individual agents and the emergent properties of
their collective behavior is analogous to those seen in many other large-scale socio-technical systems \cite{vespignani:sociotechn2012}.\\
% This is in particular the case for economies that are strongly dependent on consumer spending.\\

Accurate, valid, and timely measures of consumer confidence
are thus of pivotal importance to policy-makers and econometric forecasting. However, as a social 
and abstract construct ``consumer confidence'' is difficult to measure. Researchers have turned
to social science methods such as surveys and questionnaires which are expensive and time-consuming to conduct, and are possibly subject to
a number of personal, cultural, and social biases, e.g.~social conformity bias \cite{Nickerson1998} which will confound measures of consumer confidence
with cultural and linguistic propensities to divulge or withhold accurate information concerning one's level of confidence. The latter
also renders comparisons of consumer confidence difficult to compares across different linguistic and cultural regions.\\

Here we investigate a computational approach that leverages large-scale search engine query volumes to gauge consumer confidence.
We start from the assumption that search engine volumes reflects the issues that a population is contemporaneously pre-occupied with  \cite{Scott2014}, 
congruent with recent work in the area of market modeling \cite{Curme28072014,preis:quantifying2013}. 
Hence, consumer confidence may be manifested in the volume of certain web searches such as ``taxes'', ``investment'', and ``stocks'', but not
others, e.g.~``cloud'' and ``cat''. We focus on China since it provides an interesting case for Consumer Confidence studies given
its unique linguistic and cultural background, and the important role that the consumption patterns of its burgeoning middle-class 
are now playing in the global economy \cite{Webman2014}.\\

We obtain Google query volume time series for a number of Chinese
characters that are likely to express various facets of Chinese Consumer Confidence given their use in existing
surveys of consumer confidence in China. Using a principal component analysis, we isolate the queries that are the main indicators of 
Chinese consumer confidence \cite{Pearson1901}, and define a Chinese Consumer Confidence Index (C3I) from a linear combination
of the respective search volume data.  We cross-validate the C3I against existing gauges of consumer confidence, demonstrating
its ability to offer an accurate, timely, and informative view on consumer confidence in a 
region that has been historically underserved with regards to econometric indices.
Our results indicate that the C3I yields new information on the nature of Chinese 
Consumer Confidence. Our work may thus contribute to the science of modeling the social construct of consumer confidence and its socio-economic correlates that shape the emergent properties of economies as large-scale socio-technical systems \cite{vespignani:sociotechn2012}.

\section*{Materials and methods}
	In our investigation we rely on the following data sources:
	
	\begin{enumerate}
		\item The Chinese CCI and ECQ surveys of consumer confidence for the period under consideration.
		\item Google Trend data for a specific number of search queries corresponding to the same time period.
	\end{enumerate}
	
	Given the different nature of these surveys we use the first as an official indicator of Chinese Consumer Confidence and
	the latter to extract consumer confidence topics from which can be translated into Google queries.
	
	\subsection*{CCI consumer confidence survey}
	
	Consumer Confidence in China is mainly gauged by 2 surveys: the Consumer Confidence Index (CCI) and the 
	Economist's Confidence Questionnaire (ECQ). \\
	
\begin{figure}[h!]
		\begin{center}
			\includegraphics[width=16cm]{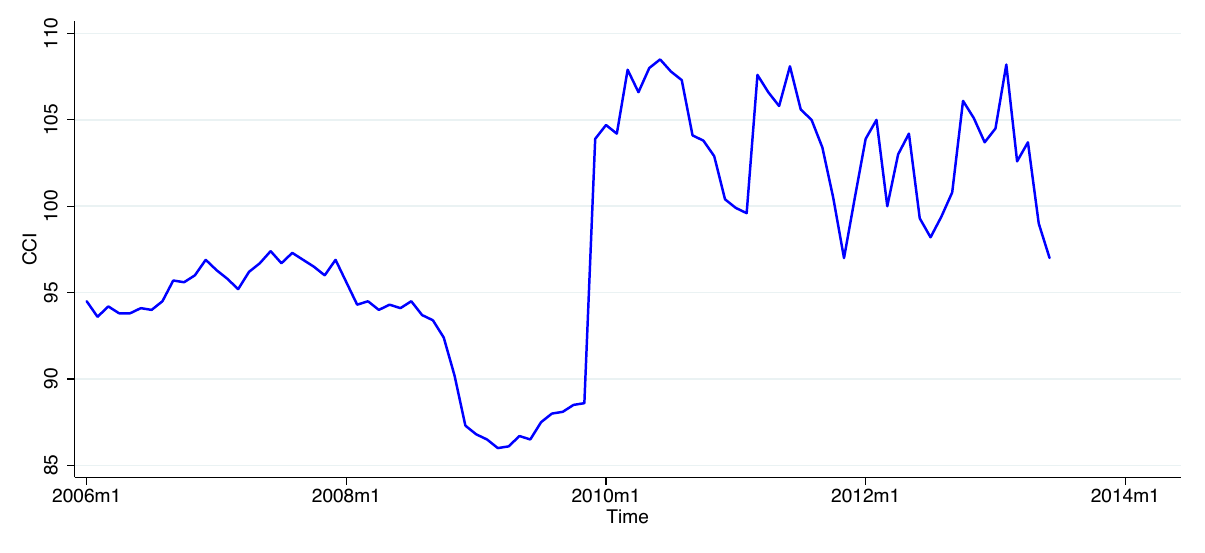}
		\end{center}
		\caption{\label{CCImonthly}Monthly time series of Chinese CCI (published by the National Bureau of Statistics of China)
							for the period of January 2006 to June 2013}
	\end{figure}

	The Chinese Consumer Confidence Index (CCI) is reported by the National Bureau of Statistics of China (NBSC) on a monthly basis. Its methodology consists
	of asking 3,500 individuals (after November, 2009) about their confidence levels of the present and the future. It consists of a questionnaire
	of about 5 simple questions each pertaining to what is assumed to be a specific component of consumer confidence, e.g. ``How do you see your current employment conditions?''.  %http://www.ce.cn/cysc/ztpd/zt/bg/
	Subjects' responses are recorded on a 5-point scale. We obtained historical monthly data of Chinese CCI from National Bureau of Statistics of China
	for the period of January 2006 to June 2013, i.e.~90 months, as shown in Fig. \ref{CCImonthly}. It must be noted that the CCI numbers 
	reported by the NBSC may be affected by some considerations with respect to data normalization and adjustments \cite{Center2010}.\\

	\subsection*{ECQ  consumer confidence topic extraction}
	The CCI is designed to be succinct and fast to administer. Hence it consists of short questions designed to be answered in terms that are directly evaluative
	of the question, e.g. ``positive'' and ``negative'' with respect to that particular question, e.g.~``How do you see your current employment conditions?''.
	However, we are looking to model the notion of Chinese Consumer Confidence as
	exhaustively as possible so we can determine its correlates in online indicators.
	
	The Economist's Confidence Questionnaire (ECQ) contains 
	31 open questions such as ``What do you presently consider the greatest threat to the Chinese economy?'', with a number
	of possible responses provided that can range from a few items to more than 15. Give the more open and exhaustive nature
	of the ECQ we manually extract the core topics of the ECQs questions and answers, and corresponding Chinese characters, to define an initial set of terms that can
	be reliably transformed to specific Google search queries. The volume of the latter are then taken to indicate the level of online attention with respect to that particular topic.\\
	
	For example, ECQ Question 13 is "How do you think the dollar value may change in the next 6 months?''. We manually extract the Chinese character for
	``dollar trend'', and add it to our initial set of topics that we deem to be indicative of consumer confidence. We then retrieve Google Trend data for each individual topic.

	As shown in Table \ref{Topics1} and \ref{ECQtopics} (Appendix), we extracted a total of 44 topics  from the ECQ questions ranging from large	macro-economic concepts
	such as ``inflation'' to more personal notions such as ``food price''. However, only 34 topics could be retained for having sufficient Google query volumes and were
	thus used as variables in our later analysis.

	\subsection*{Google Trend data}
	
\begin{figure}[h!]
		\begin{center}
			\includegraphics[width=16cm]{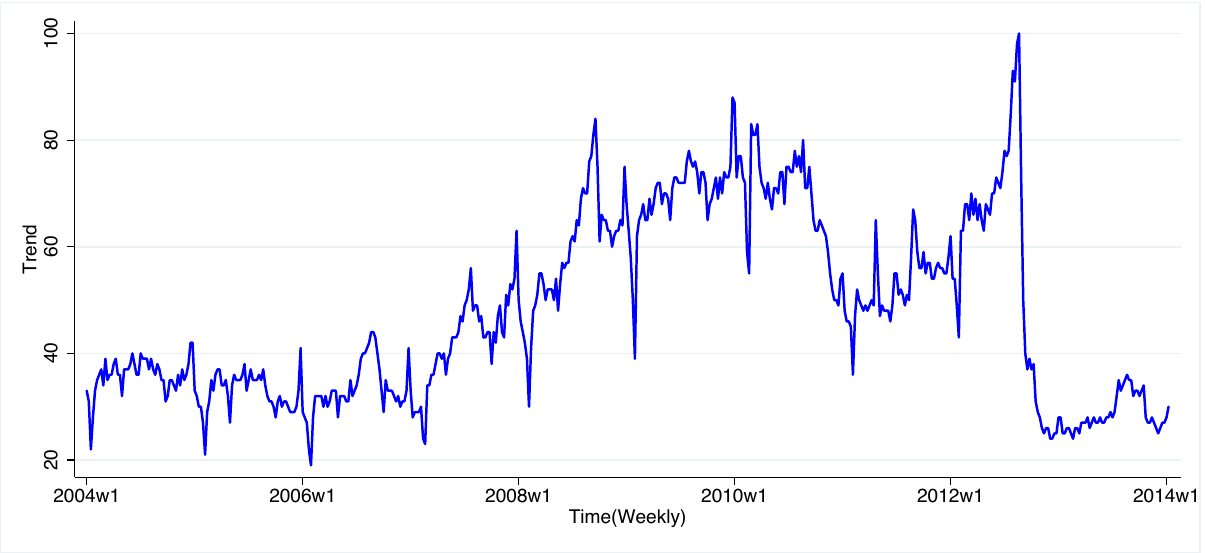}
			\caption{ \label{googletrendshoa} Google Trends graph showing weekly fluctuations of search volume for ``Hao'' (en: ``good'')}
		\end{center}
	\end{figure}

	The Google Trends (\url{www.google.com/trends/}) service is offered by the Google search engine; it allows researchers to retrieve weekly/monthly normalized search volume data
	for any user-provided search query, provided the query has non-zero search volume. 
	For example, a user can enter the query "good" and Google Trends will return a weekly time series whose
	values represent the volume of searches for that query recorded by Google in that period of time on a weekly basis.
	An example of the Google Trends data for  the Chinese character ``Hao'' (en: ``good'') is shown in Fig. \ref{googletrendshoa}.
		
	As such we obtain Google Trends data for the 34 topics that produce non-zero search volumes from January 2006 to June 2013 thereby matching the date range of our CCI data.
	Since Google Trends data can be weekly and CCI data is released monthly, we convert all weekly Google Trends time series to monthly time series by 
	means of a 4-week moving average. Since some months are longer than 4 weeks, where necessary, we move data points at the end of the month's last week to the next month.
	
	\subsection*{Methodological overview}
	
	Our research objectives are four-fold:
	
	\begin{enumerate}
		\item We model Chinese Consumer confidence from the covariances between 34 ECQ topic time series
		\item We define a Chinese Consumer Confidence Index (C3I) based on the principal components of (1)
		\item We compare our C3I to the CCI using a stepwise regression model that fits the C3I components to the CCI,
		 	including a determination of whether or not one indicator leads the other.
		\item We conduct a preliminary test of our model against a new Google Trends data set, that was not included in the original
			data that was used to construct our model (July 2013 to May 2014).
	\end{enumerate}

\begin{figure}[h!]
	\begin{center}
	      \includegraphics[width=14cm]{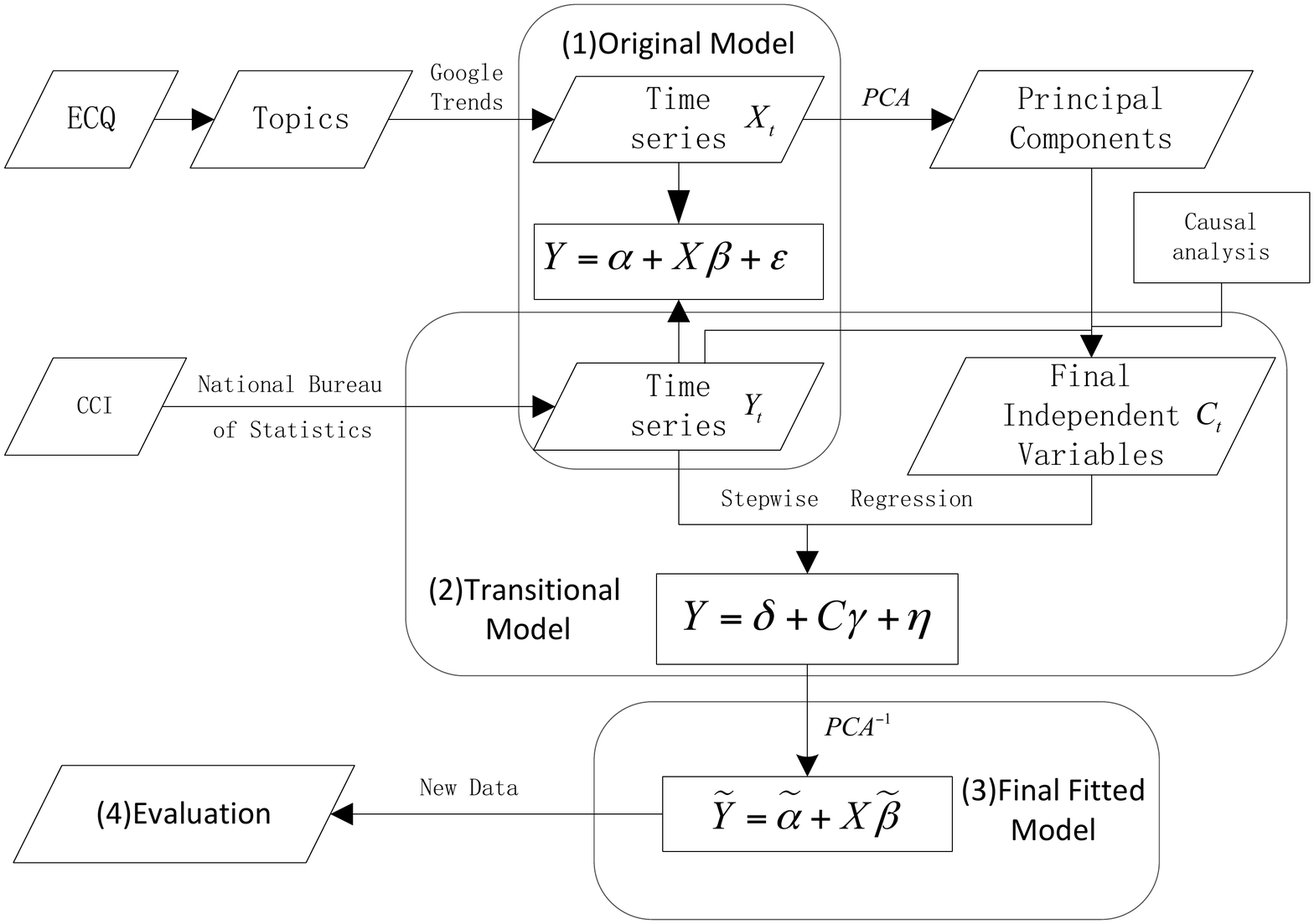}
	\caption{ \label{Methodology} Methodological overview: (a) We study the relationship between China's official CCI data ($Y$) and Google Trends Data ($X$). (b) PCA is used to determine the principal components of $X$, followed by a Granger test and VAR to determine the lead or lag relations between $X$ and $Y$. (c) $PCA^{-1}$ denotes the 
	inverse operation of PCA to obtain the fitted values of our original model. }
	\end{center}
	\end{figure}

In Fig. \ref{Methodology} we show an overview of our multi-phased methodology which is further explained in subsequent sections.

\section*{Results}

	\subsection*{Principal Component Analysis of ECQ topic covariances}

	Each of the 34 Google trends time series (corresponding to the ECQ questionnaire topics) can be taken as independent variables, 
	representing a certain facet of consumer confidence. However, we need to determine the degree of multicollinearity to investigate whether
	each variable independently represents consumer confidence, and to ensure the validity of later regression models used to fit
	a potential	 C3I based on these 34 independent variables to the CCI.\\
	
	Therefore we perform a principal components analysis (PCA) \cite{Hotelling1933} to study the components that underlie the covariances of 
	our 34 Google trends time series and reduct dimensionality. This will also ensure the orthogonality of our components and thus avoid the 
	issue of multicollinearity in 
	future regression models.\\
	
	We list the 10 highest ranked components with their loadings in Table \ref{ADFpars}. A KMO test \cite{Cureton1983} and 	
	squared multiple correlation (SMC) test\cite{Abdi2007} show that the PCA was  indeed a suitable procedure.\\
	
		\begin{table}[h!]
			\begin{minipage}[t]{.5\textwidth }
			\centering
			\vspace{0pt}
			\begin{tabular}{cll}
			\hline
			Component	&       Proportion	&	Cumulative	\\
			\hline
			1	        &       0.319              	&	0.319  \\
			2	        &       0.214	         &      0.533 \\
			3	        &       0.088	         &      0.621 \\
			4	        &       0.069	         &      0.690 \\
			5	        &       0.046	         &      0.736 \\
			6	        &       0.032	         &      0.769 \\
			7	        &       0.029	         &      0.798 \\
			8	        &       0.027	         &      0.824 \\
			9	        &       0.020	         &      0.845 \\
			10	        &       0.017	         &      0.862 \\
			\hline
			\end{tabular}
			\end{minipage}
			\begin{minipage}[t]{.5\textwidth }
			\centering
			\vspace{0pt}
			\includegraphics[width=7cm]{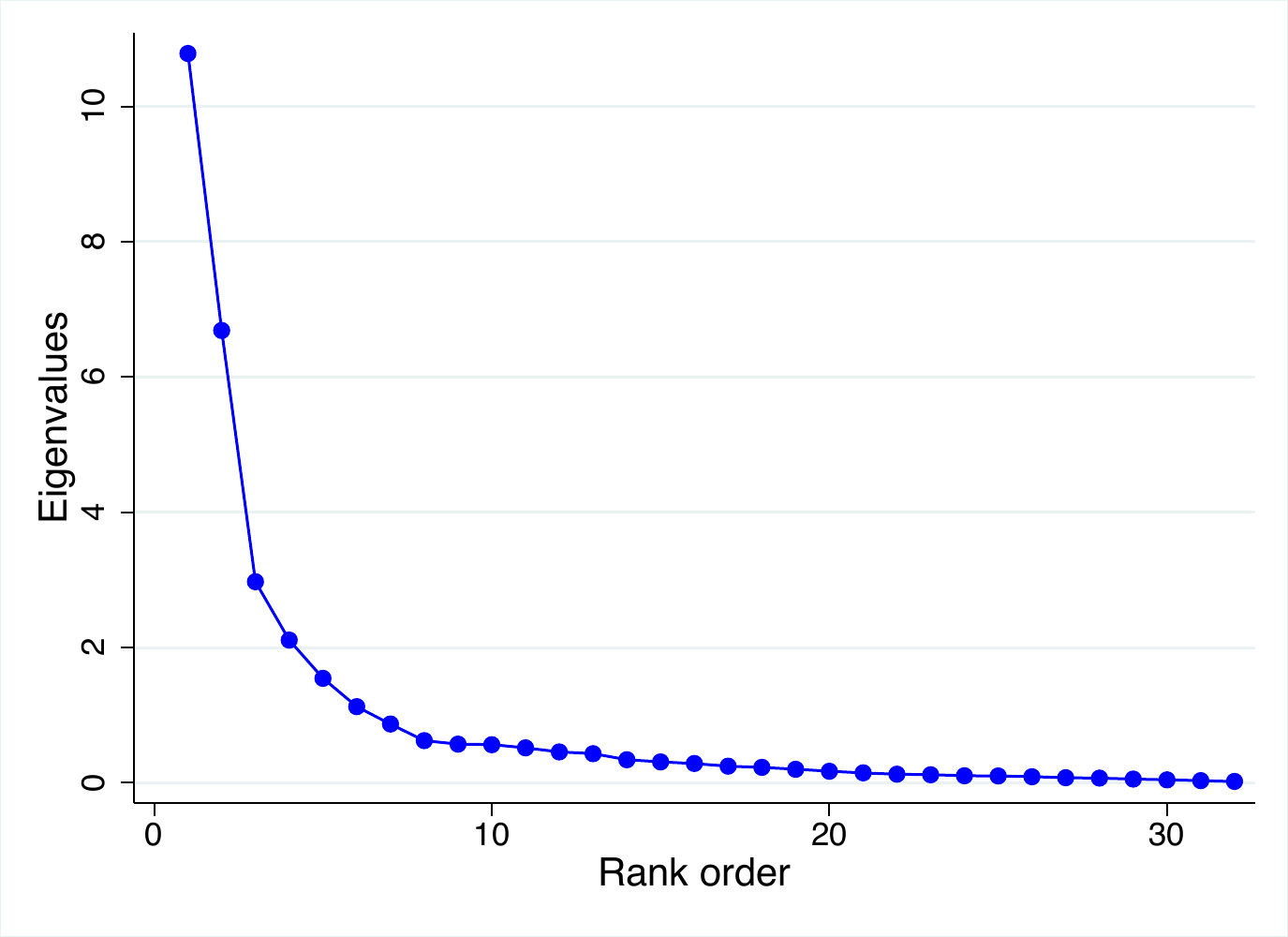}
			\end{minipage}
			\caption{\label{ADFpars} Principal Components of our 34 Google trends time series (left) with scree-plot for the eigenvalues of PCA (right). }
	\end{table}
	
	Judging from the scree-plot, we arbitrarily retain the first 9 PCA components since they represent the majority of information on the original topic covariances (about 85\%), thus ensuring we retain all relevant information for accurate modeling. However, not all 9 components need to be included in our Transitional model (Fig. \ref{Methodology}) since each carries increasingly less information. In fact, whether we choose 8, 9 or 10 components would be of little significance to our Transitional model. In fact components 7, 8 and 9 are indeed not included in some of our models below.\\
	
	We project our topic variables unto the selected 9 components only, i.e .$(C_1, C_2,...,C_9)$, i.e. we define 
	
	\[ C_{i, t}=X^T_t \times c_{i \in \{1,2,...,9\}}\]
	where $X=(x_1,x_2, ... ,x_{34})^T$ refers to our 34 topic time series and $c_i$ refers to the entries of the 9 component vectors as listed in Table \ref{comp9coef}.\\

\begin{table}
\begin{center}
\begin{tabular}{crrrrrrrrr}
\hline
			& \multicolumn{9}{c}{Components}																					\\\hline
Variables   &	$c_1$  &  $c_2$        & 	$c_3$   &    $c_4$   &	$c_5$   &	$c_6$   &	$c_7$  	&	$c_8$  &	 $c_9$	\\
\hline
$x_1$	&	-0.161	&	0.032	&	0.198	&	0.051	&	-0.034	&	-0.303	&	-0.092	&	0.494	&	-0.277	\\
$x_2$	&	-0.039	&	-0.02	0       &	0.257	&	0.337	&	-0.111	&	0.515	&	0.230	&	-0.161	&	-0.090	\\
$x_3$	&	0.141	&	-0.038	&	0.261	&	-0.307	&	0.053	&	-0.261	&	0.374	&	-0.183	&	-0.079	\\
$x_4$	&	0.243	&	-0.084	&	-0.088	&	0.121	&	-0.055	&	0.047	&	0.054	&	0.078	&	-0.008	\\
$x_5$	&	-0.037	&	0.133	&	0.335	&	0.143	&	-0.149	&	-0.187	&	0.066	&	0.254	&	0.342	\\
$x_6$	&	-0.108	&	0.004	&	0.306	&	-0.058	&	0.463	&	-0.086	&	0.234	&	0.082	&	-0.113	\\
$x_7$	&	0.242	&	-0.078	&	0.030	&	0.067	&	-0.014	&	-0.107	&	0.052	&	0.047	&	-0.130	\\
$x_8$	&	0.254	&	-0.016	&	0.129	&	0.161	&	0.088	&	0.000	&	-0.006	&	0.042	&	0.058	\\
$x_9$	&	-0.077	&	0.228	&	-0.067	&	0.238	&	0.114	&	-0.278	&	-0.048	&	-0.303	&	0.333	\\
$x_{10}$	&	0.112	&	-0.144	&	0.211	&	0.016	&	0.295	&	-0.008	&	-0.341	&	0.067	&	0.501	\\
$x_{11}$	&	0.176	&	0.033	&	0.285	&	0.077	&	0.102	&	-0.188	&	0.262	&	-0.076	&	-0.043	\\
$x_{12}$  &	0.289	&	-0.052	&	0.038	&	0.125	&	0.028	&	0.062	&	0.010	&	0.044	&	0.010	\\
$x_{13}$	 &	0.156	&	0.161	&	0.136	&	-0.32	0      &	0.016	&	0.290	&	-0.098	&	0.056	&	-0.012	\\
$x_{14}$	&	0.279	&	-0.116	&	-0.057	&	0.041	&	-0.028	&	0.054	&	0.020	&	0.030	&	-0.009	\\
$x_{15}$	&	0.070       &	0.280    	&	-0.001	&	0.030	&	-0.061	&	-0.282	&	-0.178	&	-0.275	&	-0.127	\\
$x_{16}$	  &	0.201	&	0.025	&	-0.009	&	-0.337	&	-0.017	&	-0.138	&	0.209	&	-0.294	&	0.064	\\
$x_{17}$	&	0.191	&	0.186	&	-0.110	&	0.164	&	0.270	&	-0.064	&	-0.138	&	0.042	&	-0.070     \\	
$x_{18}$	&	-0.068	&	0.292	&	0.115	&	0.050	&	0.096	&	0.286	&	-0.057	&	-0.207	&	0.100	\\
$x_{19}$	&	-0.069	&	0.247	&	0.238	&	0.218	&	-0.159	&	0.023	&	-0.036	&	0.008	&	-0.160	\\
$x_{20}$	 &	0.087	&	0.305	&	0.032	&	-0.014	&	-0.043	&	-0.081	&	-0.074	&	0.033	&	-0.140	\\
$x_{21}$	  &	0.270       &	-0.135	&	0.080	&	0.020      &	0.082	&	0.031	&	-0.100     &	0.006	&	-0.061	\\		
$x_{22}$	&	0.011	&	0.268	&	0.015	&	0.009	&	0.291	&	0.122	&	-0.187	&	-0.163	&	-0.34	0       \\
$x_{23}$	&	0.239	&	0.125	&	-0.201	&	-0.013	&	0.019	&	0.056	&	0.161	&	0.194	&	-0.054	\\
$x_{24}$	&	0.242	&	-0.177	&	0.059	&	0.013	&	-0.024	&	0.028	&	-0.158	&	0.056	&	-0.138	\\
$x_{25}$	&	-0.023	&	0.222	&	0.338	&	0.046	&	-0.183	&	0.063	&	-0.018	&	-0.045	&	-0.031	\\
$x_{26}$	&	0.101	&	0.155	&	0.148	&	-0.341	&	-0.020	&	0.086	&	-0.34	0      &	0.114	&	0.152	\\
$x_{27}$	&	0.131	&	0.212	&	-0.282	&	0.130        &	0.204	&	-0.130	&	-0.045	&	-0.034	&	-0.091	\\	
$x_{28}$	&	-0.159	&	-0.078	&	0.058	&	-0.054	&	0.545	&	0.173	&	0.079	&	0.099	&	0.002	\\
$x_{29}$	&	0.089	&	0.280	&	-0.120	&	0.109	&	-0.007	&	-0.005	&	0.139	&	0.375	&	0.082	\\
$x_{30}$	&	0.255	&	-0.061	&	0.095	&	0.107	&	0.024	&	0.044	&	-0.087	&	-0.015	&	-0.044	\\
$x_{31}$	&	-0.032	&	0.244	&	-0.245	&	-0.026	&	0.144	&	0.148	&	0.357	&	0.216	&	0.120	\\
$x_{32}$	&	0.142	&	0.227	&	-0.055	&	-0.070	&	-0.110	&	0.079	&	0.231	&	0.021	&	0.335	\\
$x_{33}$	&	0.104	&	0.209	&	0.042	&	-0.38	0       &	-0.110	&	0.123	&	-0.059	&	0.135	&	-0.074	\\
$x_{34}$	&	0.279	&	-0.003	&	0.049	&	0.188	&	-0.028	&	0.020        &	0.045	&	-0.031	&	0.033	\\	
\hline
\end{tabular}
\caption{Entries of the first 9 PCA components.}
\label{comp9coef}
\end{center}
\end{table}

To avoid spurious regression results \cite{Yule1926,Granger1974}, we must determine whether our time series have co-integrated relationships. By co-integrated relationship we refer to the possibility of a long run equilibrium relationship between two trending stationary processes which could be stationary after differencing with the same time, e.g.~$I(1)$.
Here, $I(0)$ denotes that the time series is stationary whereas $I(d)$ denotes that the time series will be stationary after $d$ times difference.   
After we extract the 9 first components, we conduct an ADF test \cite{Dickey1979,Fuller1976} to check the variables' unit root, the results of which are shown in Table \ref{ADF test}. 
The results in Table \ref{ADF test} indicate that $C_5$, $C_7$ and $C_8$ are stationary at a 1\% significance level. We define $A_{i,t}= C_{i,t}+ C_{i,t-1}, (i=5, 7, 8)$.
Subsequently all 9 variables are integrated to order one ($I(1)$) ensuring they are stationary after computing difference once. 

\begin{table}
	\begin{center}
	\begin{tabular}{cp{1.5cm}lp{1.5cm}l}
	\hline
      	Time Series	&	$x$	        &	P-Value ($x$)  &       $d.x$         & P-value ($d.x$) \\   
		 \hline
	$C_1$	&	         $I(1)$	&	0.706               &        $I(0)$	&	0.000                     \\
	$C_2$	&		$I(1)$	&	0.253               &        $I(0)$	&	0.000                     \\  
	$C_3$	&		$I(1)$	&	0.020               &        $I(0)$	&	0.000                     \\  
	$C_4$	&		$I(1)$	&	0.136               &        $I(0)$	&	0.000                     \\  
	$C_5$	&	 	$I(0)$	&	\textbf{0.000*}  &        $I(0)$ 	&	0.000                     \\ 
	$C_6$	&	  	$I(1)$	&	0.011                &        $I(0)$ 	&	0.000                     \\
	$C_7$	&		$I(0)$	&	\textbf{0.000* } &        $I(0)$ 	&	0.000                     \\
	$C_8$	&	  	$I(0)$	&	\textbf{0.007* } &        $I(0)$	&	0.000                     \\ 
	$C_9$	&		$I(1)$	&	0.023                &        $I(0)$ 	&	0.000                     \\
	  CCI        &		$I(1)$	&	0.294                &        $I(0)$	&	0.000                     \\ 
	\hline

	\end{tabular}
	\caption{\label{ADF test} Results of ADF test over first 9 PCA components, incl. the probability of having a unit root. (1)  If a variable has no unit root, then it could be seen as stationary time series. Here, $d.x$ means computing difference once to $x$. (2) P-values equaling 0 indicated $<0.0005$: we use the same notation throughout the paper.}
\end{center}

\end{table}

\subsection*{Modeling  and Computing}

After determining the principal components of our Google trend time series data, i.e.~the components that best describe consumer confidence as indicated from Google query volume with respect to our 34 survey topics, we perform a  Vector Auto-regression (VAR) \cite{Sims1980} to determine the degree of auto-correlation in our CCI data. As shown in Table \ref{VARdata}, we find a considerable degree of auto-correlation, indicating the necessity to include CCI at lag 1 as an independent variable in future analysis. This finding is intuitive, since consumers factor previous confidence into their assessment of future conditions as well as other present information.\\

We conduct a  Granger Causality test \cite{Granger1969} between our independent variables, $C_i (i=1, 2, 3, 4, 6, 9)$ and $A_j (j=5, 7, 8)$ vs. one dependent variable, namely $\mbox{CCI}_t$.
The results indicate that independent variables $C_1, C_3$ and $A_5$ are Granger causative of the $\mbox{CCI}$. Since results in behavioral science \cite{Barberis1998} indicate that people tend to discount older information in favor of newer information, we choose variables that were lagged one unit.\\

\cite{Barberis1998} indicate that people tend to discount older information in favor of newer information, we choose variables that were lagged one unit.\\
\begin{table}
\begin{center}
\begin{tabular}{clrcrc}
\hline
    Dependent Var. : CCI   &  			         &	Coef.	&	Std. Err.	&	z\   \	    &	$p>z$	\\
\hline
                             CCI      &	            L1.	&	0.735	&	0.102	&	7.190	 &	\textbf{0.000*}	\\
	                                   &	            L2.	&	0.005	&	0.109       &	0.050	 &	0.961	\\
                          $C_1$       &	            L1.	&	-0.012	&	0.021	&	-0.580	&	0.564	\\
	                                  &                  L2.       &	-0.009	&	0.021	&	-0.46	0      &	0.649	\\
                         $C_2$       &	            L1.	&	0.006	&	0.027	&	0.210	&	0.837	\\
	                                  &	            L2.	&	-0.011	&	0.026	&	-0.430	&	0.664	\\
\hline
\end{tabular}
\caption{\label{VARdata}Vector Auto-regression Results: (1) L1. (i.e. L.) indicates a 1 month time series whereas L2. indicates a 2 month lag. (2) $\mbox{L.CCI}$ is significant with a low p-value in the VAR model, so it is added in the model as an independent variables. }
\end{center}
\end{table}

\begin {table}
\begin {center}
\begin {tabular} {cllll}
\hline
Equation	&	excluded	&	$\chi^2$	&	df	&	$p>\chi^2$	\\
\hline
CCI	&	$C_1$	&	6.661	&	2	&	\textbf{0.036*}	\\
CCI	&	$C_2$	&	0.802	&	2	&	0.670	 	\\
CCI	&	$C_3$	&	6.015	&	2	&	\textbf{0.049*}	\\
CCI	&	$C_4$	&	0.980	&	2	&	0.613		\\
CCI	&	$A_5$	&	11.491	&	2	&	\textbf{0.003*}	\\
CCI	&	$C_6$	&	3.013	&	2	&	0.222		\\
CCI	&	$A_7$	&	0.371	&	2	&	0.831		\\
CCI	&	$A_8$	&	1.783	&	2	&	0.410		\\
CCI	&	$C_9$	&	0.708	&	2	&	0.702		\\
CCI	&	$ALL$     &	33.627	&	18	&	0.014		\\
\hline
\end {tabular}
\caption{ \label{Granger} Results of Granger Test. P-value marked * indicate the variable is Granger causative of the CCI at the 5\% significance level.}
\end {center}
\end {table}

The normalization of CCI data in reference to 1996 data \cite{Center2010} ended in November 2009 leading to an apparent discontinuity in the CCI data in 2009-2010 as shown in Fig.\ref{CCImonthly}.

To ensure our CCI data is not biased by structural changes, but merely a difference in normalization, we conduct as Structural Change test \cite{Pasinetti1981}. The results are summarized in Table \ref{SCT_results}; the null-hypothesis that no structural change occurred must be rejected. In other words, the results indicate a structural change is likely to have occurred in November 2009. However, it is unlikely that this change was the result of systemic changes in how consumers evaluate and express their confidence. It is rather more likely that the discontinuity results from changes in time series normalization. Hence, we do not add a dummy variable in our models to the CCI data itself but account for it elsewhere in our model.\\

\begin {table}
\begin {center}
\begin {tabular} {ll}
\hline
Structural Change Test:  $Y = X + D_0 + D_X$	&	$H_0$: no structural change	\\
\hline
Chow Test $[K, N-2*K] = 4.7362$	                         &	$p > F(13 , 62)$   0.000	\\
Wald Test = 87.3899	                                                  &	$p > \chi^2(41)$    0.000	\\
Likelihood Ratio Test = 60.6914	                                 &	$p > \chi^2(41)$    0.000	\\
\hline
\end {tabular}
\caption{\label {SCT_results}Three Structural Change Tests indicate that we can reject the hypothesis of no structural change.}
\end {center}
\end {table}

As indicated in Table \ref{SCT_results}, all three tests imply there is a structural change in the time series, which may have resulted from the NBSC standardization in November 2009.
We therefore add dummy variable D to all the independent variables of our model, with the exception of $\mbox{CCI}_{t-1}$, where the first time period comprises 47 months and the second time period comprises 42 months.\\

\begin{equation}
D_t=
\begin{cases}
  0,t\leq t_0\\
  1, t>t_0\\
\end{cases}
\end{equation}

Then, our transitional model (i.e. Model 2 in Fig. \ref{Methodology}) can be written as follows:

\begin {equation}
\mbox{CCI}_t=
\begin{cases}
 (\alpha+D_t\gamma_0) +  \sum_{\substack{i=\{1-4, 6,9\}}} (\beta_i + D_t\gamma_i) C_{i,t} + \sum_{\substack{i=\{5,7,8\}}} (\beta_i + D_t\gamma_i) A_{i,t}\\
 + \sum_{\substack{i=\{10,11\}\\
			      j=\{1,3\}}} (\beta_i + D_t \gamma_i)C_{j,t-1} + (\beta_{12} + D_t \gamma_{12})A_{5,t-1} + \delta \mbox{CCI}_{t-1}+ \epsilon_t
\end{cases}
\label{model}
\end {equation}
% \hspace{0.8cm}
 
 where $t_0=47$.\\
  
We then proceed with a Stepwise Regression \cite{Draper1998} as follows: \\
 
 \begin{enumerate}
	\item Set an appropriate significance level of 0.1.\\
	\item Fit Eq. \ref{model} by Ordinary Least Squares (OLS).\\
	\item If all the parameters pass the test, then stop, otherwise, proceed to step 4.\\
	\item Select the variable with the lowest significance level, and drop it. Fit the new equation, minus the variable, by OLS.\\
	\item Repeat step 3 and step 4, until all variables pass the test.
\end{enumerate}
 
\begin{figure}[h!]
	\begin{center}
	      \includegraphics[width=12cm]{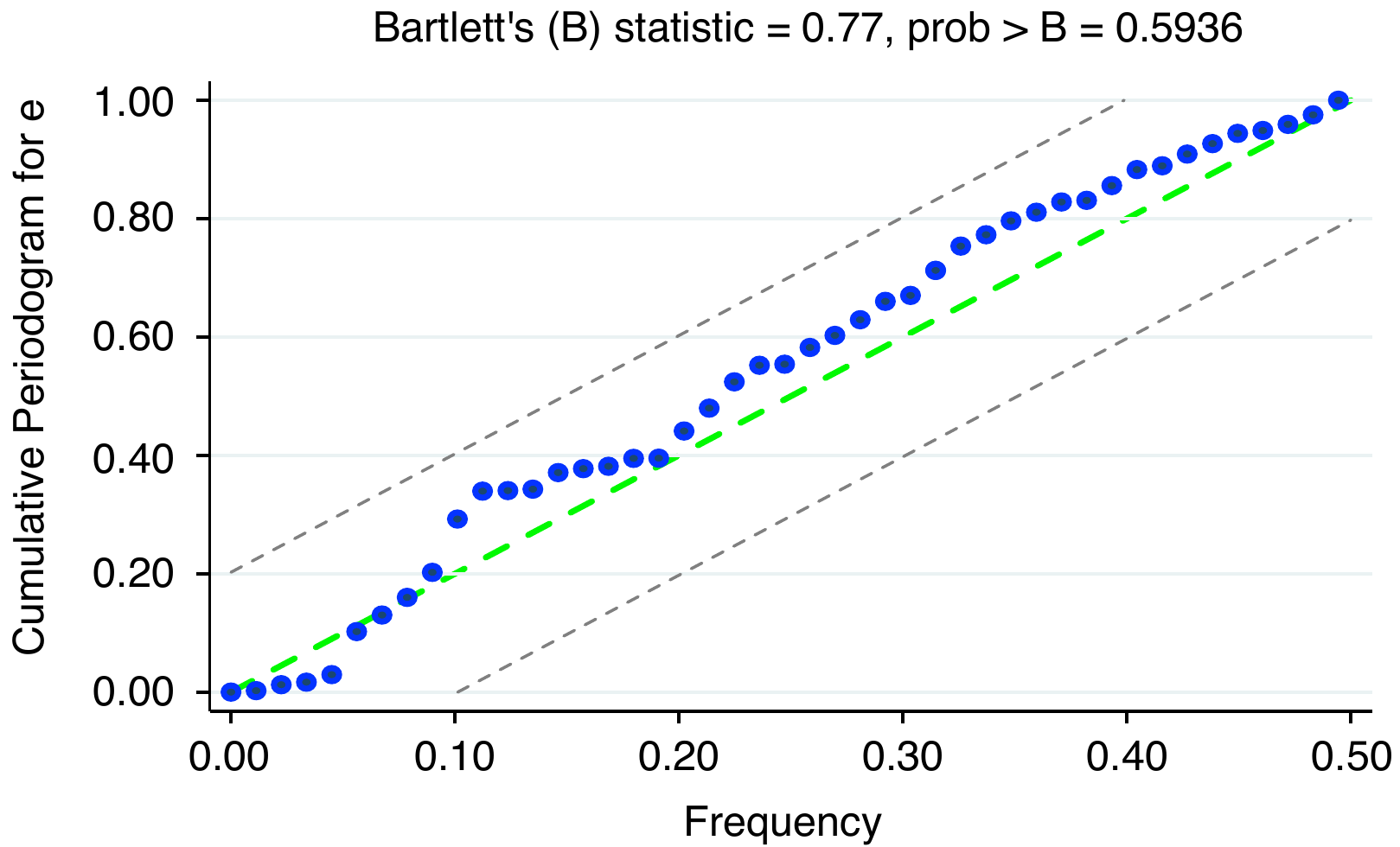}
	\caption{ \label{Bartlett} Bartlett Test for Auto-correlation indicates that our model does not exhibit auto-correlation, a necessary but not sufficient condition to ensure that the estimated parameters in classic OLS are Best Linear Unbiased Estimator (BLUE).}
	\end{center}
\end{figure}

As shown in Table \ref{RegressionResults} the model exhibits a good fit with a large $R^2$ and a small square root of residuals ($SS$). We conduct a White- and Bartlett Test \cite{Bartlett1937,White1980} to determine whether the regression has heteroscedasticity and auto-correlation. As indicated by the results shown in Table \ref{WhiteTest} and Figure \ref{Bartlett}, this is not the case.\\

 \begin {table}
 \begin {center}
	 \begin {tabular} {lrcrcl}
	 \hline
	Variables                    &	Coef.	&	Std. Error	&	 t-value	         &	$P>t$	      &  	[95\% Conf.Interval]		\\
	 \hline
	
	 $\mbox{CCI}_{t-1}$	         &	0.498	&	0.068	&	7.28	         &	0.000	       &	$[0.362, 0.634]$	\\
	$C_1$	                 &	0.025	&	0.014	&	1.88	         &	0.064	       &	$[-0.002,	  0.052]$	\\
	$C_2$	                 &	-0.023	&	0.007	&	-3.17  	&	0.002	       &	$[-0.037,	-0.009]$	\\
	$C_3$	                 &	-0.029	&	0.014	&	-2.14	         &	0.035	       &	$[-0.056, -0.002]$	\\
	$C_4$	                 &	-0.038	&	0.010	&	-3.69	         &	0.000               &	$[-0.059,	-0.018]$	\\
	 $cons$              	&	46.555      &	6.476	&	7.19	         &	0.000              &	$[33.660,	59.449]$	\\
	$dum$	                 &	10.513	&	2.304	&	4.56	        &	0.000	        &	$[5.926,	15.100]$	\\
	$du_{C_2}$	         &	-0.114	&	0.028	&	-4.05	        &	0.000	        &	$[-0.171,	-0.058]$	\\
	$du_{C_3}$	         &	0.132	&	0.028	&	4.64	        &	0.000	        &	$[0.075,	0.188]$	\\
	$du_{A_5}$	         &	-0.043	&	0.010	&	-4.25	        &	0.000	        &	$[-0.063,	-0.023]$	\\
	$du_{A_6}$	         &	0.073	&	0.017	&	4.26	        &	0.000	        &	$[0.039,	0.106]$	\\
	
	 \hline
	 $R^2=0.930$			&&	Adjust $R^2=0.920$			   &              &	$SS=1.804$				\\
	 \hline
	 \end {tabular}
	 \caption{Regression results for model represented by Eq. \ref{model}. The coefficients of $cons$ and $dum$ are denoted $\alpha$ and $\gamma_0$ whereas the coefficients of $C_i$ and $du_{C_i}$ are denoted $\beta_i$ and $\gamma_i$ respectively.}
	 \label{RegressionResults}
 \end{center}
 \end {table}

 \begin {table}
	 \begin {center}
		 \begin {tabular} {ll}
		\hline
		$H_0$: Homoskedasticity    &  \\
		\hline
		$\chi^2=51.74$                         &  $p> \chi^2 = 0.4058$\\
		\hline
		 \end {tabular}
	 \caption{ \label{WhiteTest}The results of a White Test indicate that we can not reject the hypothesis of homoscedasticity in our model. Homoscedasticity is one of necessary conditions to consider the estimated parameters from classic OLS efficient  estimators, i.e. Best Linear Unbiased Estimator property (BLUE) \cite{Gujarati2008}.}
	  \end{center}
 \end {table}

We conduct an ADF test on the residual error \cite{Dickey1991} to determine whether the regression is co-integrating or not. Table \ref{ ADF Test to $epsilon_t$} indicates that $\epsilon_t$  has no unit root, which implies that there is a co-integrated relationship in the regression, supporting the validity of our regression analysis.

 \begin {table}
 \begin {center}
 \begin {tabular} {ll}
\hline
$Z(t)=-8.321$    &    1\% critical value = -3.527\\
\hline
$p=0.0000$     &                  \\
\hline
 \end {tabular}
 \caption{The result of a  Co-integrated Relationship Test shows that there is co-integrated relationship in the regression model indicating that the regression is reasonable using time series with trend.}
 \label{ ADF Test to $epsilon_t$}
 \end{center}
 \end {table}

Using the regression results we can model C3I as  shown  in Eq. \ref{CCI_model_parameters}.

\begin {equation}
\mbox{C3I}_t =
 \begin {cases}
46.555+0.498\mbox{CCI}_{t-1}+ 0.025C_{1,t} -0.023C_2-0.030C_3-0.038C_4, t\leq 47; \\
57.067+0.498\mbox{CCI}_{t-1}+ 0.025C_{1,t}-0.137C_2+0.102C_3-0.038C_4\\
-0.043A_5+0.073C_6, t >47. 
\end{cases}
\label{CCI_model_parameters}
\end {equation}

This fitted equation preserves the major components of the PCA ($C_1-C_4$) to avoid significant information loss. We can formulate our final fitted model using the original indices as shown in Eq. \ref{fitted}.

 \begin {equation}
\mbox{C3I}_t =
 \begin {cases}

t\leq 47:	&	46.555+0.498\mbox{CCI}_{t-1}+X_t A				\\
t >47: 	&	57.067+0.498\mbox{CCI}_{t-1}+X_t B+ X_{t-1} C		\\

\end{cases}
\label{fitted}
\end {equation}

where $X^T= (x_1, x_2, ..., x_{34});$ and the entries of $A^T$, $B^T$, and $C^T$ are provided in subsequent tables and the appendix.

The  $A^T$, $B^T$, and $C^T$ matrices reveal significant changes in the structure of the C3I over time. In Tables \ref{matrixA} and \ref{matrixBC}, we show the positive and negative topics influencing our estimation of the C3I, showing how certain topics positively contribute to C3I and others contribute negatively to C3I. In particular we see that before December 2009 positive topics include "stocks", "CPI", and topics related to "trade". Negative topics notably include "prices", e.g. ``housing'', ``fuel'', ''food'',"over capacity", and concerns about "economic transition". Examining Table \ref{matrixBC} we find that these negative topics are not positive influences in C3I. In fact, the top ranked positively contributing topics are now ``over capacity'', ``real estate'', ``housing prices''. We also note that the negatively contributing topics continue to include ``exchange rates'' and ''foreign currency''.

\begin {table}
 \begin {center}
 \begin {tabular} {llll}
\hline
Positive Topics	                    &	Influence to C3I	                       &	Negative Topics	                &	Influence to C3I \\
\hline							 
Stocks	                           &	0.018	                                &	Economy transition	                 &	-0.023	\\
CPI               	                   &	0.011	                                &	Over capacity	                          &	-0.021	\\
International trade	           &	0.010	                                &	Private investment                   	&	-0.019	\\
Investment	                   &	0.010	                                &	Housing price	                          &	-0.017	\\
Trade balance	                   &	0.009	                                &	Real economy	                          &	-0.014	\\
Inflation	                           &	0.009	                                &	Dollar trend	                          &	-0.014	\\
Real estate sales	          &	0.008	                                &	Municipal bond                      	 &	-0.013	\\
Deposit reserve	                  &	0.008	                                &	Income gap	                          &	-0.010	\\
Tax	                                   &	0.007	                                &	S-M enterprise management	 &	-0.008	\\
Economic performance	 &	0.006	                                &	Exchange rate	                          &	-0.007	\\
Employment situation	 &	0.004	                                &	Interest rate for loan	                  &	-0.006	\\
GDP growth rate	         &	0.003	                                &	Food price	                          &	-0.005	\\
Demand	                          &	0.003	                                &	Crude oil price	                           &	-0.005	\\
PPI                                    	&	0.002	                                &	Employment	                           &	-0.003	\\
Foreign exchange	         &	0.002	                                 &	US economy	                          &	-0.003	\\
Fixed investment	         &	0.001	                                 &	Urbanization	                           &	-0.002	\\
	                                  &		                                         &	Consumption	                           &	-0.002	\\
	                                   &		                                          &	Population aging	                  &	-0.001	\\\hline                  
 \end {tabular}
 \caption{The Topics' Affect parameters to C3I before December 2009, i.e. matrix $A$.}
 \label{matrixA}
 \end{center}
 \end {table}

\begin {table}
 \begin {center}
 \begin {tabular} {lcc}
\hline
Topics	                         &	Influence in period $t$	   &	Influence in period $t+1$	\\
\hline
Over capacity	                &	0.057       &	0.005	\\
Real estate sales	        &	0.039	&	0.001	\\
International trade	        &	0.021	&	0.001	\\
Deposit reserve	                &	0.017	 &	0.001	\\
Housing price	                &	0.014	&	0.008	\\
Employment situation	&	0.010	&	0.001	\\
Economic performance	&	0.010	&	0.002	\\
Consumption	                 &	0.008	&	0.001	\\
CPI	                                  &	0.006	&	0.005	\\
Stocks	                          &	0.004	&	0.001	\\
Private investment	        &	0.003	&	0.006	\\
\hline
Tax	                                 &	0.032	&	-0.004	\\
Population aging	        &	0.030	&	-0.013	\\
Inflation	                         &	0.028	&	-0.001	\\
Trade balance	                 &	0.026	&	-0.002	\\
Fixed investment	        &	0.023	&	-0.001	\\
Demand	                        &	0.017	&	-0.001	\\
Employment	               &	0.012	&	-0.004	\\
M-S operation	               &	0.008	&	-0.004	\\
Income gap	               &	0.004	&	-0.020	\\
Urbanization	               &	0.004	&	-0.023	\\
\hline
Dollar trend	               &	-0.015	&	-0.004	\\
Investment	               &	-0.028	&	-0.001	\\
Exchange rate	                &	-0.039	&	-0.013	\\
PPI	                                &	-0.054	&	-0.006	\\
US economy	               &	-0.055	&	-0.012	\\
Real economy	               &	-0.074	&	-0.005	\\
Foreign exchange	       &	-0.078	&	-0.009	\\
\hline
Municipal bond	               &	-0.011	&	0.002	\\
Economy transition	       &	-0.011	&	0.007	\\
GDP growth rate	       &	-0.020	&	0.005	\\
Food price	               &	-0.040	&	0.002	\\
Crude oil price	               &	-0.053	&	0.000	\\
Interest rate for loan	       &	-0.056	&	0.003	\\

\hline
 \end {tabular}
 \caption{The Topics' Effect parameters to C3I after November 2009, matrix $B$ and $C$ indicate the influence power of $X$ to current and future C3I respectively.}
 \label{matrixBC}
 \end{center}
 \end {table}

\begin{figure}[h!]
	\begin{center}
	\includegraphics[width=15cm]{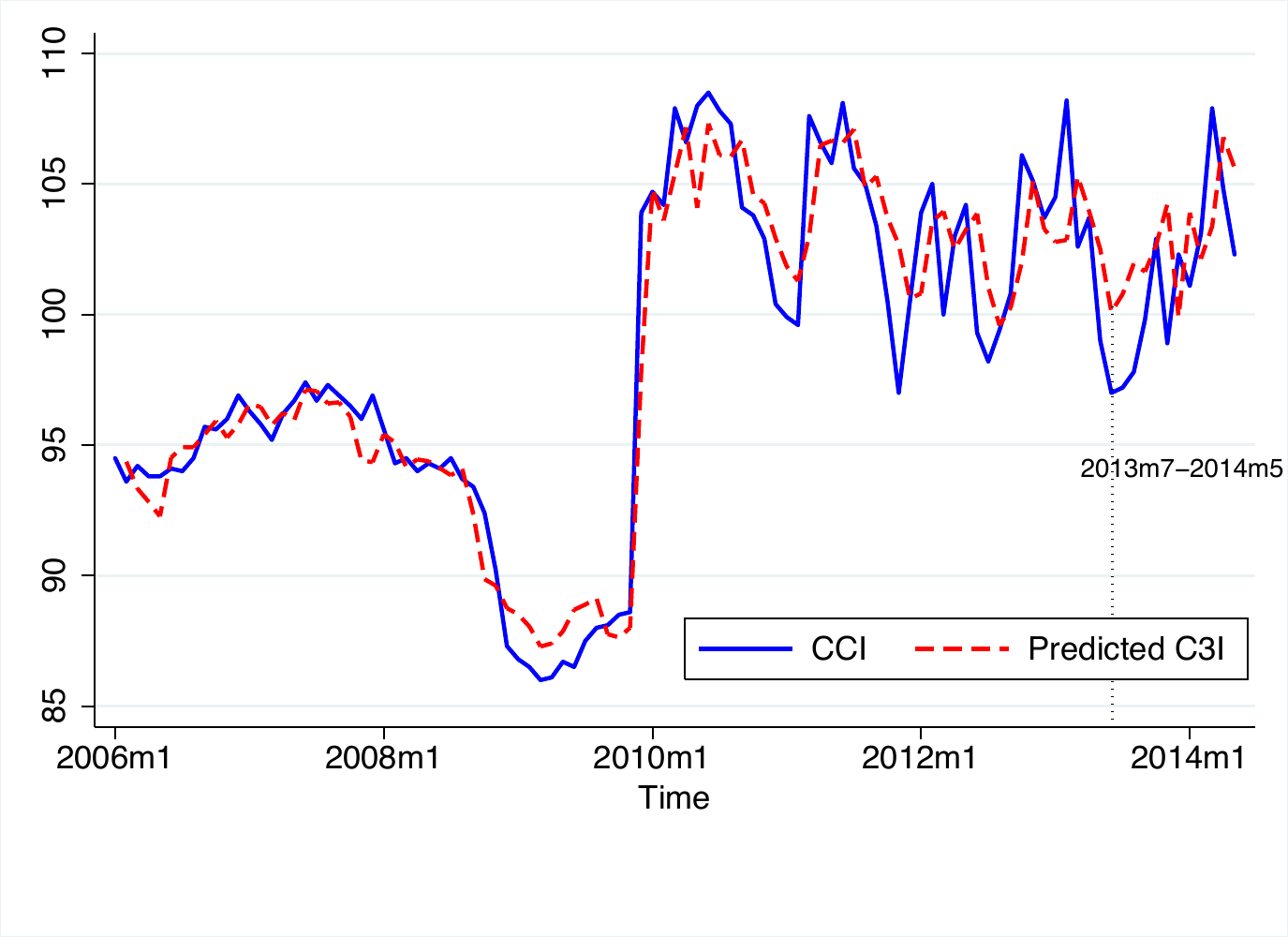}
	\caption{ \label{evaluation} Graph overlay of CCI values predicted by our C3I model vs. official CCI values reported by the National Bureau of Statistics of China. Note that the period July 2013 to May 2014 represents a prediction of CCI values on the basis of new Google trends data for that period of time as a validation of model robustness.}
	\end{center}
\end{figure}

We compare C3I values predicted by our model to the actual CCI values in Fig. \ref{evaluation} which highlights the strong degree of match between our model and actual CCI values as reported by the Chinese National Bureau of Statistics. In fact, after conducting our original analysis, we obtained new Google Trends data for the period July 2013 to May 2014, nearly a year, and re-applied the model developed from the original to this new Google Trends data. As shown in Fig. \ref{evaluation} our model predictions match the new C3I values quite well, in spite of the renormalization that Google applies to new data requests, indicating the model is robust to minor changes in the underlying Google Trends data.

\section*{Conclusions}
We model Chinese Consumer Confidence by analyzing the relationship between Chinese CCI data and Google Trends time series for queries derived from official CCI questionnaires. We subjected our Google Trends data to a PCA to reduce its dimensionality and avoid variable multicollinearity. We show the model exhibits no significant auto-correlation or heteroscedasticity using an ADF test and a White-test \cite{White1980}. Our model which includes Google Trends data as well as lagged CCI data manages to approximate the official CCI values quite well cf.~an R-square value of 0.9203, and furthermore produces a good prediction of new C3I values obtained after conclusion of the original data.\\

The resulting model allows us to draw a number of noteworthy conclusions.\\

First, our finding indicates that the results of expensive and time-consuming Consumer Confidence surveys might be approximated and potentially extended by more economical and time-efficient methods that leverage online behavioral indicators. This however requires a careful consideration of which online indicators are most relevant to the assessment of consumer confidence. Here, we focused on Google Trend data that was obtained for a narrow set of query terms (carefully derived from official Economist's Confidence Questionnaires) to ensure validity, and to avoid the introduction of noise or spurious correlations. In fact, rather than an approximation of official CCI data, the use of Google Trends data might in fact enhance the assessment of consumer confidence by avoiding structural measurement changes such as those that may have caused the discontinuity observed in the official CCI data on November 2009 leading to elevated CCI values after Nov. 2009. We were forced to introduce a dummy variable to account for this abrupt upward change of consumer confidence, which is particularly notable given the steep downward trend from 2007 and 2009.\\

Second, we observe that the C3I data is shaped by a number of inherent factors that may be fundamentally related to how people assess future economic conditions. As shown in Eq. \ref{CCI_model_parameters} we found an $\alpha=46.555$ for $t \leq 47$ and 
$\alpha + D_t \gamma_0=57.067$ for $t > 47$, as well as a $\delta$ value with respect to the lag 1 CCI data, denoted $\mbox{CCI}_{t-1}$, of 0.498. This result indicates that the C3I is partially shaped by its own previous values. We speculate that people may extrapolate their present confidence to an assessment of future economic confidence, as well as relying on other relevant information.\\

 	\begin{figure}[h!]
	\begin{center}
                \includegraphics[width=12cm]{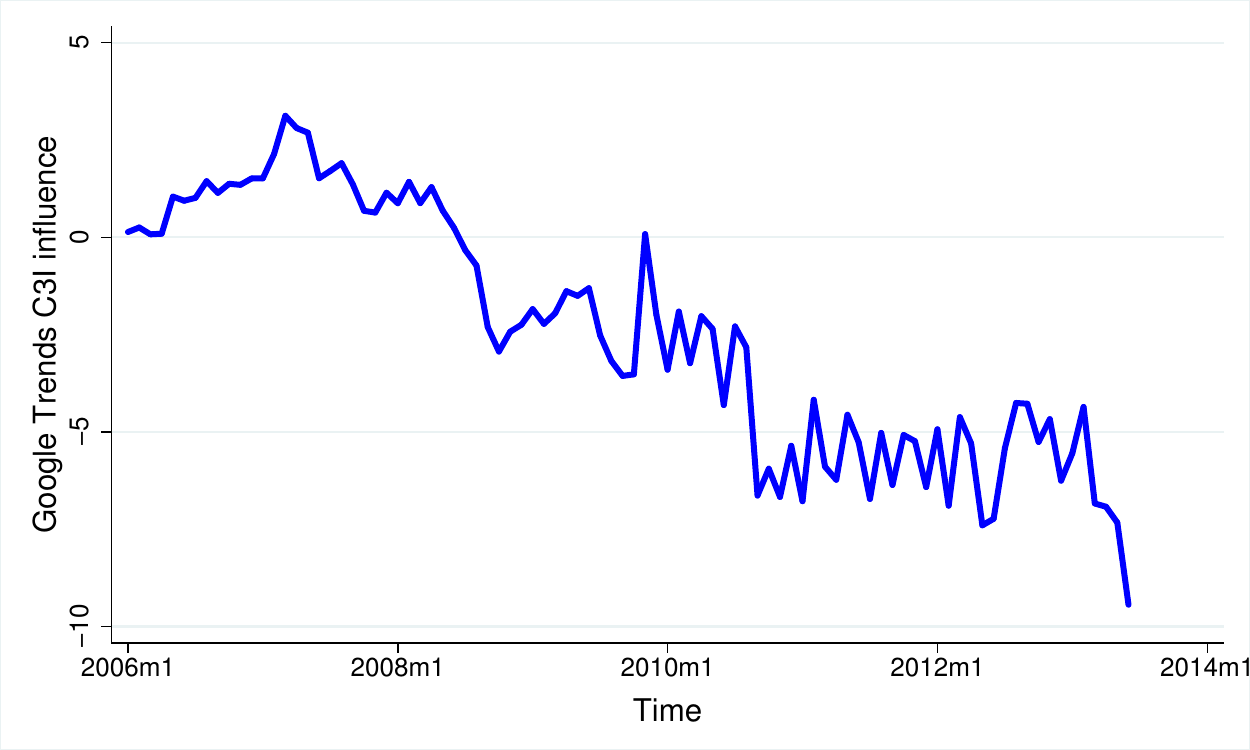}
	\caption{ \label{regressionComponent}  The contribution of Google Trends Data to our C3I model plotted over time reveals a downward trend possibly indicating that the public are losing economic confidence as judged from search engine queries.}
	\end{center}
	\end{figure}

Third, as shown in Fig. \ref{regressionComponent}, our Google Trends data indicates a consistent downtrend in consumer confidence from 2007 to the present which is not mirrored by official CCI data. However, Google Trends data presumably provides only a partial indicator of the factors that shape consumer confidence. We can therefore not conclude that our Google Trends model indicates an actual downtrend in consumer confidence. It does point to an interesting divergence between two different, but related measures of consumer confidence. We also note that after the observed discontinuity, CCI does exhibit a slight downward.\\

Fourth, examining the topics that contribute positively or negatively to our estimation of C3I reveals a number of interesting changes over time.
The first part of Eq. \ref{fitted}, i.e. $t\leq47$ corresponds to the period before December 2009. Matrix A, shown in Table \ref{matrixA}, can be split into 2 categories of topics, namely those that contribute positively to C3I and those that contribute negatively according to their coefficients. Note that the topics themselves do not contribute to C3I. The attention they receive in the population, measured by Google trends volume, is used as an indicator of the population's pre-occupation with the topic in relation to the C3I. The topics in Table \ref{matrixA} thus reveal the internal topical structure of this particular measurement of consumer confidence through a behavioral measure and which topics contribute negatively or positively to our estimation of C3I. As shown in Tables \ref{matrixA}, \ref{matrixBC} we see that a number of topics contributing positively to our estimation of C3I change polarity in C3I estimate after November 2009. This change may indicate that the population changed its assessment of these topics, leading to a different contribution to their consumer confidence, or potentially a change in how the CCI is measured. For example, when a large number of individuals search for ``over capacity'' this might occur because of the perception of over capacity as a negative issue, while some years later, people might search for the same topic from the position that over capacity is improving,  hence making a positive contribution to their consumer confidence. \\

Fifth, generally, consumer confidence can be shaped by 4 distinct consideration, namely whether one is either confident vs.~unconfident with respect to either present vs.~future conditions: one can be confident about the present and future, confident about the present but unconfident about the future, unconfident about the present but confident about the future, and, finally, unconfident about both present and future. The topic information in Table \ref{matrixBC} can express these 4 types of conditions. The table is split vertically into 4 parts, with the polarity of topics in each row alternating between positive vs.~negative for either $t$ and $t+1$. The top part of the table lists topics that have positive polarity for both $t$ and $t+1$, i.e.~these topics correspond to consumer confidence that is positive with respect to the present and future (period $t$ and $t+1$). Following we list topics that are positive with respect to time $t$ but negative with respect to $t+1$, etc. An examination of Table \ref{matrixBC} may thus reveal the degree to which certain topics contribute positively or negatively to confidence or lack therefore with respect to present and future conditions.\\

We furthermore observe that the magnitude of coefficients in the $t$ column of Table \ref{matrixBC} is generally significantly higher than that of the $t+1$ column, indicating that our topics are most suited to gauging consumer confidence about present conditions, rather than future conditions. Since the CCI is designed to measure people's confidence about current economic situations, they most likely will express their positive or negative feelings with respect to the present.  Since the polarity of topics might change over time, this might reduce their influence as indicators of future conditions.\\

In spite of the promise of this novel approach of measuring Chinese consumer confidence from search query volume, we must note a number of shortcomings that should be addressed in future work.
First, our terms and therefore topics are manually derived from the official Chinese ECQ survey. However, our choice of terms and topics might not fully capture the essence of how the CCI as a survey measures consumer confidence, since the latter consists of a different set of questions that respondents are required to answer in their entirety, not as a bag of words. Future work may focus on a more complete, principled, and thorough translation of the construction of the CCI to a set of search query terms, possible by the use of n-grams to capture more of the underlying semantics of consumer confidence.\\

Second, we assumed the CCI serves as a ground truth for measuring consumer confidence which is reasonable given it was deliberately designed to do so in terms of eliciting explicit responses. However, as a result any biases or deficiencies of the CCI will impact the validity of our own model. Furthermore, our model was optimized to match the outcomes of the CCI which may or may not in all cases accurately reflect true consumer confidence. In future work, we may include alternate gauged of consumer confidence to arrive at a more reliable and comprehensive assessment of consumer confidence to condition our model to.\\

Third, we relied on a limited, pre-defined set of topics, which may or may not provide exhaustive coverage of the notion of consumer confidence. More work should be conducted on careful selection of indices and consequently model building.\\

Fourth, our reliance on Google trends data introduces a number of issues. In particular, if a particular aspect of consumer confidence can not be gauged from search engine volume, our method won't capture it. As suggested by \cite{Lazer14032014} the validity and accuracy of our model could be improved by the inclusion of other related indicators of consumer behavior such as social media feeds, blog volume, newspaper data, etc that will allow us to reduce overfitting and noise through ``triangulation'' or ``cross-validation''. A related source of concern is that Google renormalizes their trends data continuously and may change the algorithms by which they are produced  leading to difficulties in assessing differences and significance of absolute values over time. Although variations seem to be minimal, it may render our results more difficult to replicate or reproduce. \\

Lastly, the accuracy of our model depends on the variables that we have chosen. We attempted to make reasonable choices with high face validity and attempted to avoid extraneous variable not related to consumer confidence, but can not make any definitive claims with respect to their appropriateness or completeness. Furthermore, our use of Principal Component Analysis will lead to the inclusion of topic variables that have no bearing on CCI, but can still exert a strong influence over the construction of our model. In future research, we intend to define more efficient methods for the automated selection of variables.  We caution again that similar methods have been shown to be subject to systematic challenges \cite{Lazer14032014}, which we have painstakingly sought to avoid.\\

In spite of the deficiencies of our present approach, we have demonstrated the feasibility of modeling large-scale socio-economic phenomena such as consumer confidence from behavioral online data, i.e.~Google search queries, opening new possibilities for more exhaustive, accurate, and finer-grained models of complex dynamic socio-technical systems such as a nation's economy which is shaped by the interactions of large number of autonomous agents that respond to their individual conditions as well as those of others, including global systemic information such as financial news, economic growth forecasts, GDP numbers, and inflation numbers.

\section*{Acknowledgements}

We are grateful for the support of the Chinese Scholarship Council. We thank Jingwen Li and Jun Guan for their help and early comments on our work.  We thank Giovanni Luca Ciampaglia for the insightful discussion and comments on the manuscript.

%\bibliographystyle{plain}
%\bibliography{dong_bollen_2014} %finished

\newpage

\newpage

\section*{Appendix}

	\begin{table}[ht!]
	\begin{center}
	\caption{\label{Topics1}\bf{Topics and Variables' Name}}
	\begin{tabular}{cp{5.5cm}|cp{5.5cm}}

\hline							
Variable	&	Topics	                                                                  &	Variable	&	Topics	\\
\hline							
$x_1$	&	Municipal Bond	                                                          &	$x_{23}$	&	Investment	\\
$x_2$	&	Over Capacity	                                                          &	$x_{24}$	&	Real Estate Sales	\\
$x_3$	&	Trade Balance	                                                          &	$x_{25}$	&	Housing Price	\\
$x_4$	&	Economic Perforeance                                               &	$x_{26}$	&	Deposit Reserve Rate	\\
$x_5$       &	Private Investment	                                                  &	$x_{27}$	&	Foreign Exchange	\\
$x_6$	&	Income Gap	                                                           &	$x_{28}$	&	Urbanization	\\
$x_7$	&	Employment Situation	                                          &	$x_{29}$	&	Crude Oil Price	\\
$x_8$	&	Employment	                                                           &	$x_{30}$   &	Fixed Investment	\\
$x_9$	&	Real Economy	                                                           &	$x_{31}$	&	PPI	\\
$x_{10}$	&	Population Ageing	                                                   &	$x_{32}$	&	GDP Growth Rate	\\
$x_{11}$	&	Small and Medium-sized Enterprise Management	&	$x_{33}$	&	CPI	\\
$x_{12}$	&	Demand	                                                                    &	$x_{34}$	&	Consumption	\\
$x_{13}$	&	Inflation	                                                                    &	$x_{35}$	&	Exchange Rate of Chinese Yuan Against US Dollar 	\\
$x_{14}$	&	International Trade(Import and Export)	                  &	$x_{36}$	&	Exchange Rate of Japanese Yen Against US Dollar 	\\
$x_{15}$	&	Interest Rate for Loan                                                  &     $x_{37}$	&	Real Estate Adjust	\\
$x_{16}$	&	Stocks	                                                                    &	$x_{38}$	&	Real Estate Development	\\
$x_{17}$	&	US Economy	                                                            &	$x_{39}$	&	Debt Risk	\\
$x_{18}$    &	Dollar Trend	                                                           &	$x_{40}$	&	Macro-economy 	\\
$x_{19}$	&	Economy Transition                                                     &	$x_{41}$	&	Foreign Investment	\\
$x_{20}$	&	Food Price	                                                            &	$x_{42}$	&	Administrative Expenditure	\\
$x_{21}$	&	Tax	                                                                            &	$x_{43}$	&	Investment Scale	\\
$x_{22}$	&	Exchange Rate	                                                            &	$x_{44}$	&	Labor Force	\\
\hline							
	
	\end{tabular}
	\end{center}
	
	\end{table}

	\begin{longtable}{p{1.5cm}p{13.5cm}}
		\caption{\label{ECQtopics}\bf{ECQ questions, translations, and extracted topics}}\\\hline
		Number	&	Questions 	\\\hline
		\endfirsthead
		\multicolumn{2}{c}{\tablename\ \thetable\ \textit{Continued from previous page}} \\\hline
		Number	&	Questions 	\\\hline
		\endhead
		\hline \multicolumn{2}{c}{\textit{Continued on previous page...}} \\
		\endfoot
		\endlastfoot
Q1	&	What is your judgment on the following aspects of China's economic operation? A) Macro-economy B) Demand C) Consumption	\\
Q2	&	What do you think the current situation of China's economy?	\\
Q3	&	What do you think the next six months of imports and exports growth will become?	\\
Q4	&	What do you feel the next six months, China's foreign trade balance will be?	\\
Q5	&	You expect 2013 annual GDP growth rate will be:   	\\
Q6	&	What do you consider the next six months of CPI will be? 	\\
Q7	&	What do you think the next six months of PPI will be?	\\
Q8	&	What do you think of the international crude oil and food prices over the next six months will be?	\\
Q9	&	What the current liquidity situation of the real economy is in your eyes?	\\
Q10	&	What do you think the next six months deposit reserve rate should be?	\\
Q11	&	What do you feel the next six months interest rate for loan should become?	\\
Q12	&	What do you think the current for the following currencies into RMB nominal exchange rate is in: A) Dollar B) Euro C) Pound D) Yen	\\
Q13	&	How do you think the dollar value may change in the next 6 months?	\\
Q14	&	What do you think the RMB against the U.S. dollar will become in the next six months?	\\
Q15	&	What do you think the foreign exchange balance of China will be in the next six months?	\\
Q16	&	What do you think the domestic stock market prices in the next three months will be?	\\
Q17	&	What do you think the growth of direct foreign investment will be?	\\
Q18	&	How do you think China's fixed asset investment growth will be? What would be the growth rate of whole year?	\\
Q19	&	You believe the investment in real estate development in 2013 will increase ?  over the same period of the previous year.	\\
Q20	&	How do you think the trend of housing price for the next six months will be?	\\
Q21	&	How do you think the real estate sales over the next six months will be? 	\\
Q22	&	What?s your idea concerning ? next six months the trend of the U.S. economy? 	\\
Q23	&	What do you think the European debt crisis situation will be in the next six months?	\\
Q24	&	How do you think the exchange rate of Japanese Yen against US dollar in the next six months will be? 	\\
Q25	&	How do you think the employment situation of China this year ?	\\
Q26	&	In response to the declining trend to the labor force,  what measures can be taken in your opinion? 	\\
	&	( 1 ) to promote agricultural moderate scale of operations, improve the efficiency of land productivity	\\
	&	( 2 ) to increase investment in education , improve population quality	\\
	&	( 3 ) to encourage various technological innovation,and improve total factor productivity	\\
	&	( 4 ) to intensify reform efforts to further reform the bonus release	\\
	&	( 5 ) to make the family planning policy to respond appropriately adjust to demographic changes	\\
	&	( 6 ) to moderate lower economic growth	\\
	&	( 7 ) to accelerate the pace of economic restructuring	\\
	&	( 8 ) Others ( please specify ) :	\\
Q27	&	Comparing with the last 6 months, what do you think the local government funding in the next half year?	\\
Q28	&	What's your judgment to local government debt risk? 	\\
Q29	&	Aiming at the local fiscal revenue growth becoming slow, the cumulative risk of local financing becoming heavy, what measures do you think should be taken? 	\\
	&	( 1 ) to determine the reasonable revenue growth target	\\
	&	( 2 ) to make a reduction of administrative expenses 	\\
	&	( 3 ) to bind government investment to related entities	\\
	&	( 4 ) to expand the participation of private investment 	\\
	&	( 5 ) to improve the tax system , increase local property rights	\\
	&	( 6 ) to open up new sources of revenue ( such as resource tax, housing property tax , etc. )	\\
	&	( 7 ) to give the local government permission of issued debt	\\
	&	( 8 ) to improve income distribution mechanism of state-owned monopoly	\\
	&	( 9 ) to appropriate to relax regulation of real estate	\\
	&	( 10 ) Others ( please specify ) :	\\
Q30	&	What is the biggest risk facing the Chinese economy in 2013 in your opinion? 	\\
	&	( 1 ) Inflation	\\
	&	( 2 ) decline in economic growth	\\
	&	( 3 ) the slow progress in economy transition	\\
	&	( 4 ) the blind expansion of investment	\\
	&	( 5 ) ignore the quality in the urbanization process	\\
	&	( 6 ) significant fluctuations in housing prices	\\
	&	( 7 ) labor shortages	\\
	&	( 8 ) the risk of local financing	\\
	&	( 9 ) private lending risk	\\
	&	( 10 ) Exacerbated by overcapacity	\\
	&	( 11 ) Larger pressure on energy saving and emission reduction	\\
	&	( 12 ) the degradation of export 	\\
	&	( 13 ) ineffective environmental resources protection 	\\
	&	( 14 ) further deterioration to the income gap	\\
	&	( 15 ) operational difficulties for small and medium-sized enterprise	\\
	&	( 16 ) Others( please specify ) :	\\
Q31	&	31. What is your suggestion to the further macroeconomic policies and reforms?	\\
\hline
		
	\end{longtable}

\end{document}